.

# Modeling the Thermal Extraction of Water Ice from Regolith

P.T. Metzger[1]

[1] Florida Space Institute, University of Central Florida, 12354 Research Parkway, Partnership 1 Building, Suite 214, Orlando, FL 32826-0650; PH (407) 823-5540; email: philip.metzger@ucf.edu

## ABSTRACT

Modeling has been developed to support the development of volatile extraction technologies for the Moon, Mars, asteroids, or other bodies. This type of modeling capability is important to avoid the high cost of multiple test campaigns in simulated lunar conditions as the hardware design is iterated. The modeling uses the Crank-Nicholson algorithm applied in a two dimensional (2D) axisymmetric (extendable to 3D) finite difference formalism. It uses soil thermal parameters developed from Apollo soil measurements with adaptations for asteroid regolith. Simulations show that it successfully replicates thermal measurements on the surfaces of asteroids and the Moon and helps interpret those measurements to provide insight into the subsurface properties of those bodies. The 2D simulations have provided insight into the cooling of a lunar drill bit and provide a method to determine the original subsurface temperature despite the presence of the warm bit.

## EXTRATERRESTRIAL WATER

NASA's Lunar CRater Observation and Sensing Satellite (LCROSS) impact into a permanently shadowed crater in the Moon's south polar region kicked up an ejecta blanket that contained water, carbon monoxide, carbon dioxide, and other volatiles (Colaprete et al., 2010; Gladstone et al., 2010), definitively showing there is water on the Moon. Carbonaceous asteroids contain hydroxylated phyllosilicate minerals (Jewitt et al., 2007) that release water when heated (Zacny et al., 2016). Mars has glacial deposits of ice, hydrated mineral deposits, and water adsorbed onto the regolith grains (Abbud et al., 2016). The water and hydroxyl on these bodies can be extracted for use in support of exploration missions to lower the mission's cost and expand its capabilities (Sanders, et al., 2008).

## WATER EXTRACTION BY HEATING

Several groups have performed water extraction test with icy regolith simulant or with asteroid simulant containing hydroxylated and hydrated minerals (See e.g., Metzger et al., 2011). Tests by Zacny et al. (2016) showed that the water moves away from a heating device in the soil, so if the geometry of the mining device fails to contain the vapor or fails to drive it in the correct direction, the vapor may diffuse away through the regolith and be re-frozen somewhere else. Design of a mining device must take this into account. Testing these systems will be challenging because

.

it will require large icy-regolith beds in vacuum conditions (Kleinhenz and Linne, 2013). Optimizing the mining systems can be accelerated if we have high-fidelity modeling capable of handling all the relevant physics.

## 1D THERMAL MODEL

The first step in developing such a model was to write a 1-dimensional heat transfer model based on the Crank-Nicolson algorithm because it is unconditionally stable. The heat transfer equation is

$$\rho C \frac{\partial T}{\partial t} = k \nabla^2 T + \dot{q} \tag{1}$$

where $t$ is time, $\rho = \rho(z)$ is the material density assumed unchanging in time, $z$ is depth below the surface, $C = C(\rho, T)$ is the heat capacity, $k = k(\rho, T)$ is thermal conductivity, $T=T(z,t)$ is temperature, and $\dot{q} = \dot{q}(z,t)$ is volumetric heat source. Reducing the Laplacian to one dimension and discretizing in the forward Euler method, the equation is

$$\rho_i C_i^n \frac{(T_i^{n+1} - T_i^n)}{\Delta t} = \frac{k_{i-}^n T_{i-1}^n - (k_{i-}^n + k_{i+}^n) T_i^n + k_{i+}^n T_{i+1}^n}{(\Delta z)^2} + \dot{q}_i \tag{2}$$

in the $n$th time step at spatial location indexed by $i$, where $k_{i-}^n = (k_i^n + k_{i-1}^n)/2$ and $k_{i+}^n = (k_i^n + k_{i+1}^n)/2$ are the thermal conductivities in each direction. Alternatively, this could be written in the backward Euler method,

$$\rho_i C_i^n \frac{(T_i^{n+1} - T_i^n)}{\Delta t} = \frac{k_{i-}^n T_{i-1}^{n+1} - (k_{i-}^n + k_{i+}^n) T_i^{n+1} + k_{i+}^n T_{i+1}^{n+1}}{(\Delta z)^2} + \dot{q}_i \tag{3}$$

where for solvability we have kept $k_i^n$ in the $n$th time step. This introduces a tiny time lag in the soil properties, which becomes negligible with sufficiently small time step. Averaging the two representations gives the Crank-Nicolson method, which has the advantage of unconditional stability,

$$\frac{2(\Delta x)^2 \rho_i C_i^n}{\Delta t} (T_i^{n+1} - T_i^n) = k_{i-}^n (T_{i-1}^n + T_{i-1}^{n+1}) - (k_{i-}^n + k_{i+}^n)(T_i^n + T_i^{n+1})$$
$$+ k_{i+}^n (T_{i+1}^n + T_{i+1}^{n+1}) + 2(\Delta z)^2 \dot{q}_i \tag{4}$$

Defining

$$\alpha_{i+}^n = \frac{k_{i+}^n}{2\rho_i C_i^n} \frac{\Delta t}{(\Delta z)^2} \quad, \quad \alpha_{i-}^n = \frac{k_{i-}^n}{2\rho_i C_i^n} \frac{\Delta t}{(\Delta z)^2} \tag{5}$$

and moving the $n+1$ terms to the left side and the $n$ terms to the right,

.

$$-\alpha_{i-}^n T_{i-1}^{n+1} + (1 + \alpha_{i+}^n + \alpha_{i-}^n) T_i^{n+1} - \alpha_{i+}^n T_{i+1}^{n+1} =$$
$$\alpha_{i-}^n T_{i-1}^n + (1 - \alpha_{i+}^n - \alpha_{i-}^n) T_i^n + \alpha_{i+}^n T_{i+1}^n + 2(\Delta z)^2 \dot{q}_i^n \tag{6}$$

This takes the form of a matrix equation with temperature vectors $\hat{T}$, tridiagonal matrices $A$ and $B$, and heat source vector $\hat{\dot{q}}$,

$$A\hat{T}^{n+1} = B\hat{T}^n + \hat{\dot{q}}^n \tag{7}$$

which can be solved efficiently on a computer using the tridiagonal matrix algorithm given initial conditions and the source vector inputs.

During the lunar day, the upper layer of regolith is nearly in radiative equilibrium with insolation, so the energy that conducts beneath the surface is the tiny difference between two large values, introducing a computational problem. This work follows Vasavada et al. (1999) in using quadratic interpolation for the upper temperatures. Here, this is implemented assuming there is a radiatively-dominated layer of thickness $\Delta z$ with temperature $T_0^n$ at its center but $T_{skin}^n$ at its upper surface defined in each time step by exact radiative equilibrium. Between $T_{skin}^n$ and $T_0^n$ there is $\Delta z/2$ spacing. $T_0^n$ is solved through the quadratic interpolation between $T_{skin}^n$, $T_1^n$, and $T_2^n$, while the latter two are solved iteratively via the matrix equations. $T_0^n$ appears in the matrix equations in the source term $\dot{q}_1^n$. For layers $i = 2$ through $N-1$, $\dot{q}_i^n = 0$, but for $i=N$ the source term is equal to the geothermal heat flux from below. This geothermal flux is non-zero on the Moon as determined by the Apollo heat flow experiments, but it should be zero for asteroids. For insolation, the model tracks the sun angle during the body's rotation. The model does not included finer details such as the body's libration (Moon) or nutation/precession (asteroids), although these can be added, and it assumes constant solar flux $\psi = 1360.8$ W/m² for the Moon and can be adjusted for other bodies. Following the analysis of Diviner data (Vasavada et al. 2012) that follows the form of Keihm (1984), albedo in the lunar case is directional

$$a = 0.1 + 0.045 \left(\frac{\omega}{\pi/4}\right)^3 + 0.14 \left(\frac{\omega}{\pi/2}\right)^8 \tag{8}$$

where $\omega$ is the sun angle, $\cos\omega = |\cos\phi \cos\theta|$, and $\theta$ and $\phi$ are the sun's longitude and latitude relative to the soil as the body rotates. Emissivity is set to 0.978 near the value of Vasavada et al. (1999) and Vasavada et al. (2012). These albedo and emissivity models have also been adopted for the asteroid case, but additional work is needed to constrain them.

Parameterizing the model's soil properties relies upon published measurements in the literature, but these are incomplete especially for the case of asteroids so the following adaptations were made. Bulk density $\rho_i$ varies on the Moon with location and depth beneath the surface, but it is not well known for asteroids. Lunar values are chosen consistent with Apollo core tubes and other Apollo measurements. Asteroid bulk densities are typically determined by fitting the results of thermal modeling to the observed thermal inertias of the asteroids. Measurements by Rosetta during flyby

.

of 21 Lutetia indicates the thermal inertia increases below the top few centimeters "in a manner very similar to that of Earth's Moon" (Keihm et al., 2012). This could indicate particle sizing and/or bulk density variations over that depth, but apart from this very little is known of possible vertical structure in asteroid regolith. The parameters in this model can be adjusted to match future spacecraft measurements to help determine asteroid regolith structure.

Heat capacity and thermal conductivity for the lunar case were measured experimentally with actual lunar soil and found to be strong functions of temperature. Here the model uses the thermal conductivity parameters from the Diviner data analysis by Vasavada et al., (2012), based upon the functional form by Mitchell and dePater (1994), but modified here to a functional form allowing random, non-monotonic variation in soil density with depth,

$$k_{\text{Lunar}} = k_S + \frac{(k_d - k_S)\rho}{(\rho_d - \rho_s)}\left(1 + \chi \left(\frac{T}{350}\right)^3\right) \qquad (9)$$

where $\rho$ is bulk density of the soil, $k_S = 0.0006$ W/m/K is the conductance term for loose surficial soil at $\rho_s = 1300$ kg/m³, $k_d = 0.007$ W/m/K is the conductance term for dense soil at $\rho_d = 1800$ kg/m³, $\chi = 2.7$ scales the internal (pore space) radiative contribution, and $T$ is in kelvins. For preliminary assessment of asteroid cases, this model was modified by a factor for particle sizing, since asteroid surfaces are expected to be much coarser than the lunar surface,

$$k_{\text{Asteroid}} = \left[k_S + \frac{(k_d - k_S)\rho}{(\rho_d - \rho_s)}\left(1 + \chi \left(\frac{T}{350}\right)^3\right)\right] \times \left(\frac{D}{50\ \mu\text{m}}\right)^{1/2} \qquad (10)$$

This factor uses a power law derived by the measurements of Presley and Christensen (1997, see Table 3), who found the power law varies with atmospheric pressure. Below about 5 torr pressure the power index becomes ½. A curve was fit to their data showing that this power law is valid all the way to zero pressure. This factor uses 50 μm in the denominator because this is the approximate mass-averaged particle size of average lunar soil that determined $k_{\text{Lunar}}$.

For specific heat, the model uses the equation by Hemingway, Robie and Wilson (1973),

$$C(T) = -23.173 + 2.127\ T + 0.01501\ T^2 - 7.3699 \times 10^{-5} T^3 + 9.6552 \times 10^{-8} T^5 \qquad (11)$$

in J/kg/K. This is a fit to measurements on Apollo soil samples 14163, 15301, 60601, and 10084 to represent the average of lunar soil, and it fits the data by better than 10% over the relevant temperature range. In future work, this will be amended to vary proportionally to the soil's bulk density, so values will change by ±16% around the mid-value. For the asteroid case this equation is adopted but with the bulk density scaling implemented to explore possible asteroid surface conditions.

.

Simulations have been performed for various lunar latitudes and asteroid cases using a variety of soil density profiles, which will be reported in a future publication. An example of a 1D simulation for the lunar case is shown in Fig. 1, showing similar form to the simulations by others. An example for the asteroid case is shown in Fig. 2 for asteroid 101955 Bennu. The resulting range of temperatures matches well the range observed on Bennu per Lauretta et al. (2015).

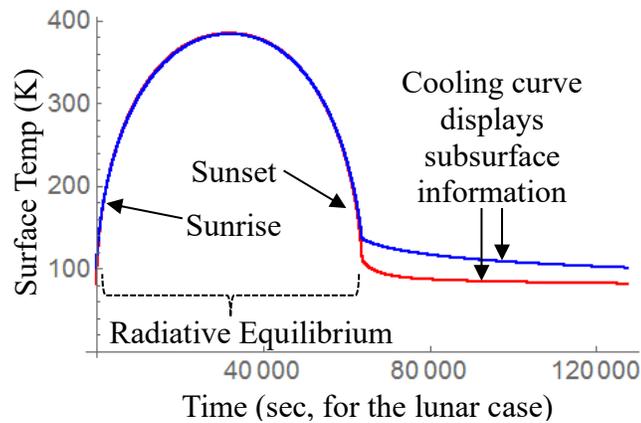

**Figure 1.** 1D modeling of the lunar case. Blue: a dense layer located 1 cm below the surface. Red: a dense layer located 5 cm below the surface.

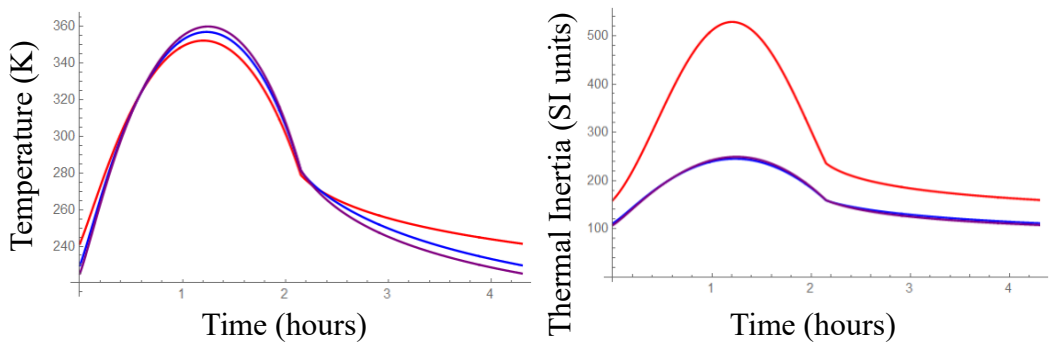

**Figure 2.** 1D modeling of asteroid 101955 Bennu near its equator. Red: homogenous regolith (no subsurface structure) with 60 μm mean particle size. Blue and purple: a surface lag of 1.5 cm gravel at the top of 60 μm sized fines deflated 5 mm and 2 mm, respectively, below the top of the gravel.

## 2D AXISYMMETRIC THERMAL MODEL

The 1D model must be converted to 2D axisymmetric to model drilling technologies. The derivation is shown in Appendix A. The result is a system of two equations that utilizes an inter-time-step temperature field $T_{ij}^*$,

.

$$(1 + \delta_z^2)T_{ij}^* = (1 - \delta_r^2)T_{ij}^n$$
$$(1 + \delta_r^2)T_{ij}^{n+1} = (1 - \delta_z^2)T_{ij}^* \qquad (12)$$

where $\delta_z^2$ and $\delta_r^2$ are the appropriate second derivative operators defined in the appendix. This can be solved efficiently using the tridiagonal matrix algorithm.

Using the same lunar soil property equations as in the 1D model, an example of this 2D model solved numerically is shown in Figure 3. This is for a drilling test in which a warm drill bit is embedded in soil that carries away its heat. Radiative heat transfer occurs between the upper surface of the soil and the walls of the surrounding vacuum chamber, while the cylindrical soil container is chilled at constant temperature and conductively removes heat from the soil.

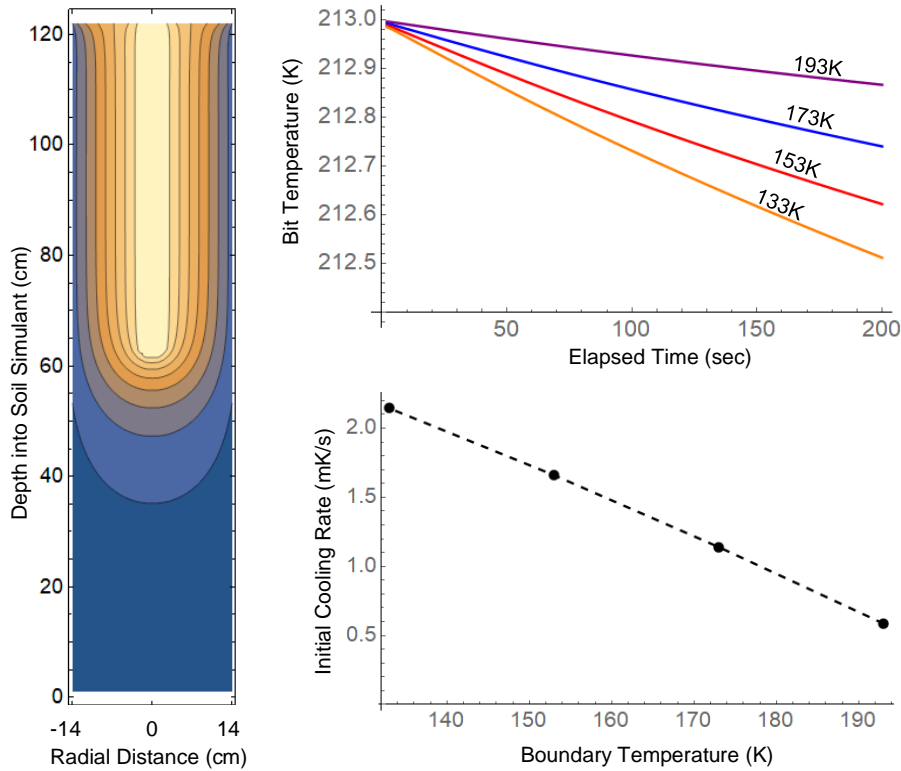

**Figure 3.** Left: 2D axisymmetric simulation of warm drill bit in a cylinder of lunar simulant. Yellow is bit temperature, 213K; Blue is container temperature, 173K for this case. Right top: Cooling curves for four cases with different container temperatures but the same initial bit temperature. Right bottom: Initial cooling rate of the bit for these four cases correlated to container temperature.

The simulations show the cooling process is so slow in lunar soil that it takes hours for soil around a warm drill bit to return to ambient temperature. It is therefore not practical for a lunar rover to sit long enough for the soil to cool to get an undisturbed subsurface temperature measurement after drilling. However, the model can predict

.

the cooling rate of the bit as a function of the original subsurface soil temperature before drilling (i.e., the boundary value at large radius). Thus, the lunar drill bit needs only measure its cooling rate (in the range of millikelvins/s) long enough for small measurement uncertainty. Then, comparing the measurement with this type of modeling will recover the original subsurface temperature. This demonstrates the usefulness of this type of modeling.

**FUTURE WORK**

Next steps include incorporating phase change of the mixed ices in the lunar permanently shadowed regions and including gas diffusion via the Darcy, Knudsen, and transitional regimes. Laboratory measurements are needed to calibrate material parameters for ice/regolith mixtures and for a wider range of asteroid materials. Then, drilling tests should be performed to include phase change of the ice to benchmark the model. The model will be extended to Cartesian 3D similar to the first step in Appendix A but with the extra dimension. Fully 3D modeling may be needed because the ice may be blocky chunks that are not axisymmetric around the drill. Finally, the model will be used to numerically experiment with different drilling and ice extraction methods for use in the lunar permanently shadowed craters or to study heating and volatile release in asteroids.

**CONCLUSIONS**

Thermal modeling has been successfully developed for the 1D and axisymmetric 2D cases for regolith in vacuum conditions. The modeling can handle both lunar and asteroid cases. For the latter case, the modeling will help constrain parameter values by comparing with spacecraft data, while in the lunar case the parameter values are already constrained by laboratory measurements of returned lunar samples, so the modeling is at a higher level of fidelity. Future work will add phase change and diffusion of volatiles. The modeling will make it possible to numerically investigate different volatile extraction methods on the Moon and asteroids.

**APPENDIX A: DERIVATION OF 2D AXISYMMETRIC EQUATIONS**

Ignoring the source heating term for now, the 2D heat flux equation in Cartesian coordinates assuming $\Delta r = \Delta z$ and using Crank-Nicolson formalism is straightforwardly

$$\frac{2(\Delta z)^2 \rho_i C_i^n}{\Delta t}\left(T_{ij}^{n+1} - T_{ij}^n\right) =$$
$$k_{i-,j}^n\left(T_{i-1,j}^n + T_{i-1,j}^{n+1}\right) - \left(k_{i-,j}^n + k_{i+,j}^n\right)\left(T_{i,j}^n + T_{i,j}^{n+1}\right)$$
$$+ k_{i+,j}^n\left(T_{i+1,j}^n + T_{i+1,j}^{n+1}\right) + k_{i,j-}^n\left(T_{i,j-1}^n + T_{i,j-1}^{n+1}\right)$$
$$- \left(k_{i,j-}^n + k_{i,j+}^n\right)\left(T_{i,j}^n + T_{i,j}^{n+1}\right) + k_{i,j+}^n\left(T_{i,j+1}^n + T_{i,j+1}^{n+1}\right) \quad \text{(A1)}$$

Converting this to axisymmetric form requires only the last line be modified for the derivative in the radial coordinate because

.

$$\nabla_z^2 T = \frac{\partial^2}{\partial^2 z} T \quad \text{vs.} \quad \nabla_r^2 T = \frac{\partial^2}{\partial^2 r} T + \frac{1}{r}\frac{\partial}{\partial r} T \tag{A2}$$

so the additional term is needed. Therefore, in cylindrical coordinates,

$$\frac{2(\Delta z)^2 \rho_i C_i^n}{\Delta t}(T_{ij}^{n+1} - T_{ij}^n) =$$
$$k_{i-,j}^n(T_{i-1,j}^n + T_{i-1,j}^{n+1}) - (k_{i-,j}^n + k_{i+,j}^n)(T_{i,j}^n + T_{i,j}^{n+1}) + k_{i+,j}^n(T_{i+1,j}^n + T_{i+1,j}^{n+1})$$
$$+ k_{i,j-}^n(T_{i,j-1}^n + T_{i,j-1}^{n+1}) - (k_{i,j-}^n + k_{i,j+}^n)(T_{i,j}^n + T_{i,j}^{n+1}) + k_{i,j+}^n(T_{i,j+1}^n + T_{i,j+1}^{n+1})$$
$$+ k_{i,j}^n[(T_{i,j+1}^n + T_{i,j+1}^{n+1}) - (T_{i,j-1}^n + T_{i,j-1}^{n+1})]/2j \tag{A3}$$

where $r_j = j\Delta r\ (= j\Delta z)$. Collecting terms as before,

$$(1 + \alpha_{i-,j}^n + \alpha_{i+,j}^n + \alpha_{i,j-}^n + \alpha_{i,j+}^n)T_{ij}^{n+1} - \alpha_{i+,j}^n T_{i+1,j}^{n+1} - \alpha_{i-,j}^n T_{i-1,j}^{n+1}$$
$$- \left(\alpha_{i,j-}^n - \frac{\alpha_{i,j}^n}{2j}\right)T_{i,j-1}^{n+1} - \left(\alpha_{i,j+}^n + \frac{\alpha_{i,j}^n}{2j}\right)T_{i,j+1}^{n+1} =$$
$$(1 - \alpha_{i-,j}^n - \alpha_{i+,j}^n - \alpha_{i,j-}^n - \alpha_{i,j+}^n)T_{ij}^n + \alpha_{i+,j}^n T_{i+1,j}^n + \alpha_{i-,j}^n T_{i-1,j}^n$$
$$+ \left(\alpha_{i,j-}^n - \frac{\alpha_{i,j}^n}{2j}\right)T_{i,j-1}^n + \left(\alpha_{i,j+}^n + \frac{\alpha_{i,j}^n}{2j}\right)T_{i,j+1}^n \tag{A4}$$

This is no longer a simple tridiagonal matrix because it indexes in both $i$ and $j$, but it can be handled by adapting the method of Summers (2012) to the axisymmetric case. First, we define operators

$$\delta_z^2 T_{ij} = -\alpha_{i-,j} T_{i-1,j} + (\alpha_{i-,j} + \alpha_{i+,j})T_{i,j} - \alpha_{i+,j} T_{i+1,j} \tag{A5}$$

and

$$\delta_r^2 T_{ij} = \left(\alpha_{i,j-} - \frac{\alpha_{i,j}}{2j}\right)T_{i,j-1} - (\alpha_{i,j-} + \alpha_{i,j+})T_{i,j} + \left(\alpha_{i,j+} + \frac{\alpha_{i,j}}{2j}\right)T_{i,j+1} \tag{A6}$$

so the equation becomes

$$(1 + \delta_z^2 + \delta_r^2)T_{ij}^{n+1} = (1 - \delta_z^2 - \delta_r^2)T_{ij}^n \tag{A7}$$

where again for solvability we kept the indices for $\alpha$ at $n$ instead of $n+1$. In a well-behaved system the fourth order cross-derivatives are very small, and their changes in time are even smaller,

$$\delta_z^2 \delta_r^2 T_{ij}^{n+1} - \delta_z^2 \delta_r^2 T_{ij}^n \approx 0 \tag{A8}$$

Subtracting this negligible difference from the equation, rearranging, and factoring,

$$(1 + \delta_z^2 + \delta_r^2)T_{ij}^{n+1} = (1 - \delta_z^2 - \delta_r^2)T_{ij}^n - (\delta_z^2 \delta_r^2 T_{ij}^{n+1} - \delta_z^2 \delta_r^2 T_{ij}^n)$$

.

$$(1 + \delta_z^2 + \delta_r^2 + \delta_z^2\delta_r^2)T_{ij}^{n+1} = (1 - \delta_z^2 - \delta_r^2 + \delta_z^2\delta_r^2)T_{ij}^n$$

$$(1 + \delta_z^2)(1 + \delta_r^2)T_{ij}^{n+1} = (1 - \delta_z^2)(1 - \delta_r^2)T_{ij}^n \tag{A9}$$

We next define $T_{ij}^*$ to within a constant of integration,

$$(1 + \delta_z^2)T_{ij}^* = (1 - \delta_r^2)T_{ij}^n \tag{A10}$$

and we substitute this into the right hand side of our equation,

$$(1 + \delta_z^2)(1 + \delta_r^2)T_{ij}^{n+1} = (1 - \delta_z^2)(1 + \delta_z^2)T_{ij}^* \tag{A11}$$

Commuting the two parentheses on the right hand side, since the derivatives are commuting operators,

$$(1 + \delta_z^2)(1 + \delta_r^2)T_{ij}^{n+1} = (1 + \delta_z^2)(1 - \delta_z^2)T_{ij}^* \tag{A12}$$

The terms operated upon by the $(1 + \delta_z^2)$ operators on each side of the equation must therefore be equal to within a constant of integration, so equating them only determines the constant of integration and thereby completes the definition of $T_{ij}^*$,

$$(1 + \delta_r^2)T_{ij}^{n+1} = (1 - \delta_z^2)T_{ij}^* \tag{A13}$$

We now have a system of two equations,

$$\begin{aligned}(1 + \delta_z^2)T_{ij}^* &= (1 - \delta_r^2)T_{ij}^n \\ (1 + \delta_r^2)T_{ij}^{n+1} &= (1 - \delta_z^2)T_{ij}^*\end{aligned} \tag{A14}$$

which recovers tridiagonal forms on both the left and right sides of each. However, each equation operates upon a different dimension (*z* or *r*) on the left and right sides, so the rows and columns are transposed and they are not representable as matrix equations. Nevertheless, they can be solved numerically the same as if they were by using the tridiagonal matrix algorithm on each piece. Thus, it is efficient to solve.

It should be noted that the *j*=0 cells on the centerline of the coordinate system need special treatment. This was done straightforwardly using the discretization of Scott and Ko (1968), carrying it consistently through the above derivation.

## ACKNOWLEDGEMENT

This work was directly supported by NASA's Solar System Exploration Research Virtual Institute cooperative agreement award NNA14AB05A. Another portion of this work was directly supported by subcontract to Honeybee Robotics on NASA SBIR contract no. NNX15CK13P, "The World is Not Enough (WINE): Harvesting

.

Local Resources for Eternal Exploration of Space." Another portion was supported by the University of Central Florida's Florida Space Institute.

.

.